\documentclass[12pt]{article}
\usepackage{epsfig}\parskip 5pt plus 1pt
\usepackage{amssymb}
\newcommand{\be}{\begin{equation}}
\newcommand{\ee}{\end{equation}}
\newcommand{\bea}{\begin{eqnarray}}
\newcommand{\eea}{\end{eqnarray}}

\def\simge{\mathrel{%
   \rlap{\raise 0.511ex \hbox{$>$}}{\lower 0.511ex \hbox{$\sim$}}}}
\def\simle{\mathrel{
   \rlap{\raise 0.511ex \hbox{$<$}}{\lower 0.511ex \hbox{$\sim$}}}}

\begin{document}
\thispagestyle{empty}
\vspace*{1cm}
\begin{center}
{\Large{\bf A novel technique for the measurement of the electron neutrino cross section} }\\

\vspace{.5cm}
A.~Longhin$^{\rm a}$, L.~Ludovici$^{\rm b}$, F.~Terranova$^{\rm c}$  \\
\vspace*{1cm}
$^{\rm a}$ I.N.F.N., Laboratori Nazionali di Frascati,
Frascati (Rome), Italy \\
$^{\rm b}$ I.N.F.N., Sezione di Roma, Rome, Italy \\
$^{\rm c}$ Dep. of Physics, Univ. of Milano-Bicocca and INFN, Sezione di Milano-Bicocca, 
Milano, Italy \\
\end{center}

\vspace{.3cm}
\begin{abstract}
\noindent
Absolute neutrino cross section measurements are presently limited by
uncertainties on $\nu$ fluxes. In this paper, we propose a technique
that is based on the reconstruction of large angle positrons in the
decay tunnel to identify three-body semileptonic $K^+ \rightarrow e^+
\pi^0 \nu_e$ decays. This tagging facility operated in positron
counting mode (``event count mode'') can be employed to determine the
absolute $\nu_e$ flux at the neutrino detector with ${\cal O}(1\%)$
precision. Facilities operated in ``event by event tag mode''
i.e. tagged neutrino beams that exploit the time coincidence of the
positron at source and the $\nu_e$ interaction at the detector, are
also discussed.
\end{abstract}

\newpage

\section{Introduction}
\label{introduction}

A detailed knowledge of neutrino interaction cross sections plays a
crucial role in the precision era of oscillation
physics~\cite{Formaggio:2013kya,Alvarez-Ruso:2014bla}. In the last
decade, a vigorous experimental programme has been pursued, employing
both the near detectors of running long-baseline
experiments~\cite{Gran:2006jn,Adamson:2009ju,Dobson:2013uxa,Abe:2013jth,Abe:2014nox}
and dedicated
experiments~\cite{Nakajima:2010fp,Tice:2014pgu,Acciarri:2014isz} with
special targets and PID capabilities. The large statistics accumulated
so far and the careful strategy implemented for systematic mitigation
have improved our knowledge of total and differential cross sections
for $\nu_\mu$ and $\bar{\nu}_\mu$ in the range of interest (0.3-5 GeV)
for future long-baseline and sterile neutrino
experiments~\cite{review_xsect}. All these experiments are, however,
designed to work in $\nu_e$ appearance mode and the direct measurement
of $\nu_e$ interactions still relies on scarce
data~\cite{Blietschau:1977mu,Abe:2014agb}.  Calculations are thus
based on extrapolation from $\nu_\mu$ results.  Despite lepton
universality of weak interactions, the ratio between $\nu_\mu$ and
$\nu_e$ suffers from uncertainties due to nuclear
effects~\cite{Day:2012gb} that have to be constrained with data to
reduce systematic errors in future long baseline $\nu_e$ appearance
experiments~\cite{Dusini:2012vc,Coloma:2012ji}.  To cope with this
challenge, novel experimental approaches have been proposed with the
aim of producing pure, intense and well controlled sources of electron
neutrinos~\cite{Volpe:2003fi,McLaughlin:2004va,Oldeman:2009wa,Adey:2013pio,Spitz:2014hwa}. The
technique proposed in the following has a similar aim: electron
neutrinos are produced by the three body decay of $K^+$ ($K_{e3}$,
i.e. $K^+ \rightarrow e^+ \nu_e \pi^0$) in standard neutrino beams.
The positrons are identified in the decay tunnel by purely
calorimetric techniques and the beam-line is optimized to enhance the
$\nu_e$ components from $K_{e3}$ and suppress to a negligible level
the $\nu_e$ contamination from muon decays.  This approach - from here
on called ``event count mode'' - has several advantages. It provides a
source of electron neutrinos that can be used to study $\nu_e$
interactions in a direct manner, i.e. without relying on
extrapolations from $\nu_\mu$. In addition, it delivers an observable
(the positron rate) that can be directly linked to the rate of $\nu_e$
at the far detector through the three body kinematics of $K_{e3}$. The
positron rate in the decay tunnel thus determines the flux with a
precision significantly better than what is currently achieved with
conventional untagged $\nu_\mu$ beams ($\sim 10\%$).  Finally, this
facility paves the way for the realization of tagged neutrino
beams~\cite{hand1969,Pontecorvo:1979zh,denisov,bernstein,Ludovici:1996sx}
in the configuration proposed in Ref.~\cite{Ludovici:2010ci}, where
the positron is associated to the corresponding $\nu_e$ interaction at
the far detector on an event by event basis (``event by event tag
mode'').  In this mode, full kinematic reconstruction of the $K_{e3}$
can be achieved measuring the photon pair from $\pi^0$ decay, thus
retrieving information on the energy of $\nu_e$ for each tagged event.

The tagging concept and the rationale for the choice of the beam-line
parameters, the tagging detector and the neutrino detector are
introduced in Sec.~\ref{sec:concept}. The beam-line up to the decay
tunnel is detailed in Sec.~\ref{sec:beamline} together with the
expected secondary flux ($\pi$ and $K$) at CERN, Fermilab, JPARC and
Protvino.  The decay tunnel instrumented with positron taggers and the
corresponding positron identification performance are summarized in
Sec.~\ref{sec:decaytunnel}. This section also summarizes the rates and
integrated doses expected at the tagger units. Background, systematics
and rates at the far detector are presented in Sec.~\ref{sec:backgr}
and Sec.~\ref{sec:rate}. Finally, perspectives for the event by event
tag mode upgrade are described in Sec.~\ref{sec:double_tag}.

\section{Conceptual design}
\label{sec:concept}

Unlike neutrino factories~\cite{Geer:1997iz} and beta
beams~\cite{Zucchelli:2002sa}, conventional neutrino beams are
sources of muon neutrinos from pion decays, polluted 
by small fractions of electron neutrinos from kaons and muons decays.
The size of the contamination highly depends on the primary
proton energy, on the momentum of secondaries selected by the focusing
system and on the length of the decay tunnel. In general, high energy
neutrino beams as the CNGS~\cite{Baldy:1999dc} are contaminated by
$\nu_e$ originating from the $K_{e3}$ decays of $K^+$ while the
contamination of lower energy neutrino beams is mostly due to $\pi^+
\rightarrow \mu^+ \nu_\mu \rightarrow e^+ \nu_e \bar{\nu}_\mu \nu_\mu$.
The $\nu_e$ flux depends on the hadron production yield on the
target and on the acceptance of the focusing and transport system to
the decay tunnel. Even with dedicated hadro-production data, 
pion monitoring at the target and muon
monitoring at the beam dump, the uncertainty on the size of this
contamination has never been reduced below 10\%. 
It is a fair educated guess that for a conventional facility, a
dedicated effort, including ancillary experiments to measure the kaon
production rate in replica targets, might reduce this uncertainty to a
level not lower than $7-8\% $.

\begin{figure}
\centering\includegraphics[width=\textwidth]{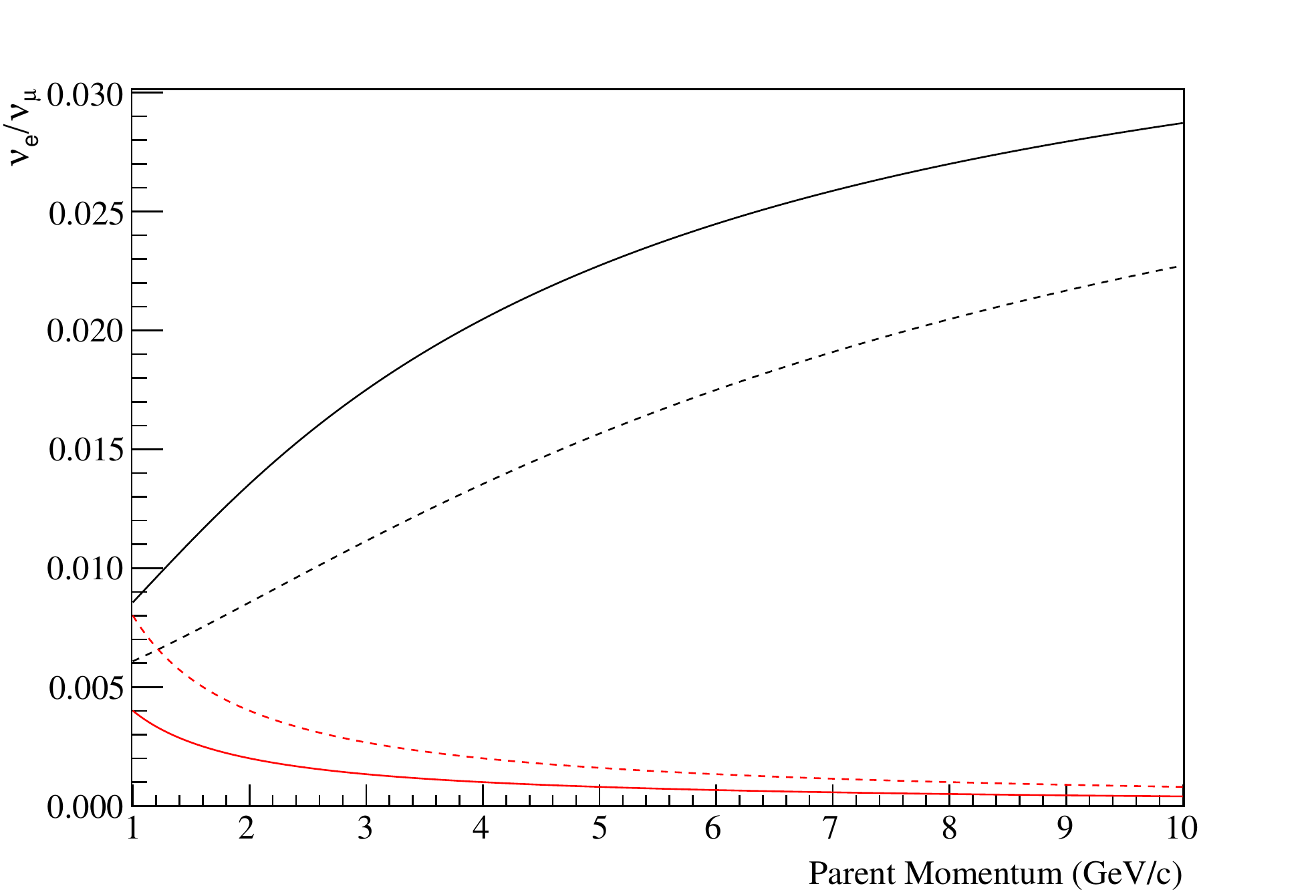} 
\caption{Black lines: approximate scaling (see Eq.~\ref{eq:scaling}) of the
  $\nu_e/\nu_\mu$ fluxes as a function of the momentum of
  secondaries. The continuous (dashed) line corresponds to a 50~m (100~m) decay
  tunnel. The red lines show the approximate scaling of the $\nu_e/\nu_\mu$ from 
muon DIF.}
\label{fig:scaling}
\end{figure}

The ratio between the $\nu_e$ from $K_{e3}$ and the $\nu_\mu$ from
pion decay can be enhanced increasing the energy of the selected
secondaries and reducing the length of the decay tunnel
(Fig.~\ref{fig:scaling} - black lines). This comes at the expenses of
the overall neutrino flux. The $\nu_e$ beam contamination from muon
decays in flight (DIF) is also reduced (Fig.~\ref{fig:scaling} - red
lines). The $\nu_e/\nu_\mu$ ratio scales as
\be
R_{K/\pi} \cdot  BR(K_{e3}) \cdot 
\frac{ \left[ 1- e^{-L/\gamma_K c \tau_K} \right]}
{\left[ 1- e^{-L/\gamma_\pi c \tau_\pi} \right]} 
\label{eq:scaling}
\ee
where $R_{K/\pi}$ is the ratio between $K^+$ and $\pi^+$ produced at
the target and transported down to the entrance of the decay
tunnel. $BR(K_{e3})$ is the $K_{e3}$ branching ratio: $5.07 \pm 0.04 \
\%$~\cite{PDG}. $L$ is the length of the decay tunnel. $\tau_K$
($\tau_\pi$) and $\gamma_K$ ($\gamma_\pi$) are the lifetime and
Lorentz factor of the $K^+$ ($\pi^+$), respectively.  
The scaling
of Eq.~\ref{eq:scaling} is depicted in Fig.~\ref{fig:scaling} assuming
$R_{K/\pi} = 10\%$ (see~Tab.\ref{tab:yield-horn} below).

For a beam dominated by the $K_{e3}$ contamination, the only source of
primary positrons in the decay tunnel is the $K^+ \rightarrow \pi^0
e^+ \nu_e$ decay and the Dalitz from the $\pi^0 \rightarrow e^+ e^-
\gamma$ decay (BR $\simeq 1.2 \ \%$). All other positrons are either
due to DIF of muons (from pions or beam halo) or to photon conversions
in the material around the decay tunnel. Two body positron decays
($\pi^+ \rightarrow e^+ \nu_e$ and $K^+ \rightarrow e^+ \nu_e$) are
chirality-suppressed and can be neglected. As a consequence, all
primary positrons are originated by three-body decays and are
distributed at angles much larger than the angles of the muons from
two-body $\pi^+ \rightarrow \mu^+ \nu_\mu$ decays. For the beam
parameters considered in Sec.~\ref{sec:beamline}, the mean positron
angle (88 mrad) is 22 times larger than the corresponding mean $\mu^+$
angle and $\sim 30$ times larger than the beam divergence of the undecayed
particles. These considerations~\cite{Ludovici:2010ci} support the
instrumentation of the decay tunnel with detectors having a geometry
 similar to the calorimeters of hadron colliders (hollow
cylinders). As discussed in Sec.~\ref{sec:decaytunnel}, the technology
requirements (radiation hardness, fast readout, fast recovery time for
pile-up mitigation etc.)  are quite similar, too. Since neither the
muons from $\pi^+$ decay nor the bulk of undecayed particles cross the
calorimeter before reaching the beam dump, the particle rate is much
smaller than the rate of muon monitors in conventional neutrino beams;
such rate (see below) can be handled by standard detector and readout
technologies developed for the hadron colliders.

Finally, the need for short decay tunnels reduces the size of the
calorimeter and makes the instrumentation of the whole tunnel - which
was considered far-fetched in 1979~\cite{Pontecorvo:1979zh} - a viable
option.

\section{Production and transport of secondaries}
\label{sec:beamline}

\begin{figure}
\centering\includegraphics[width=\textwidth]{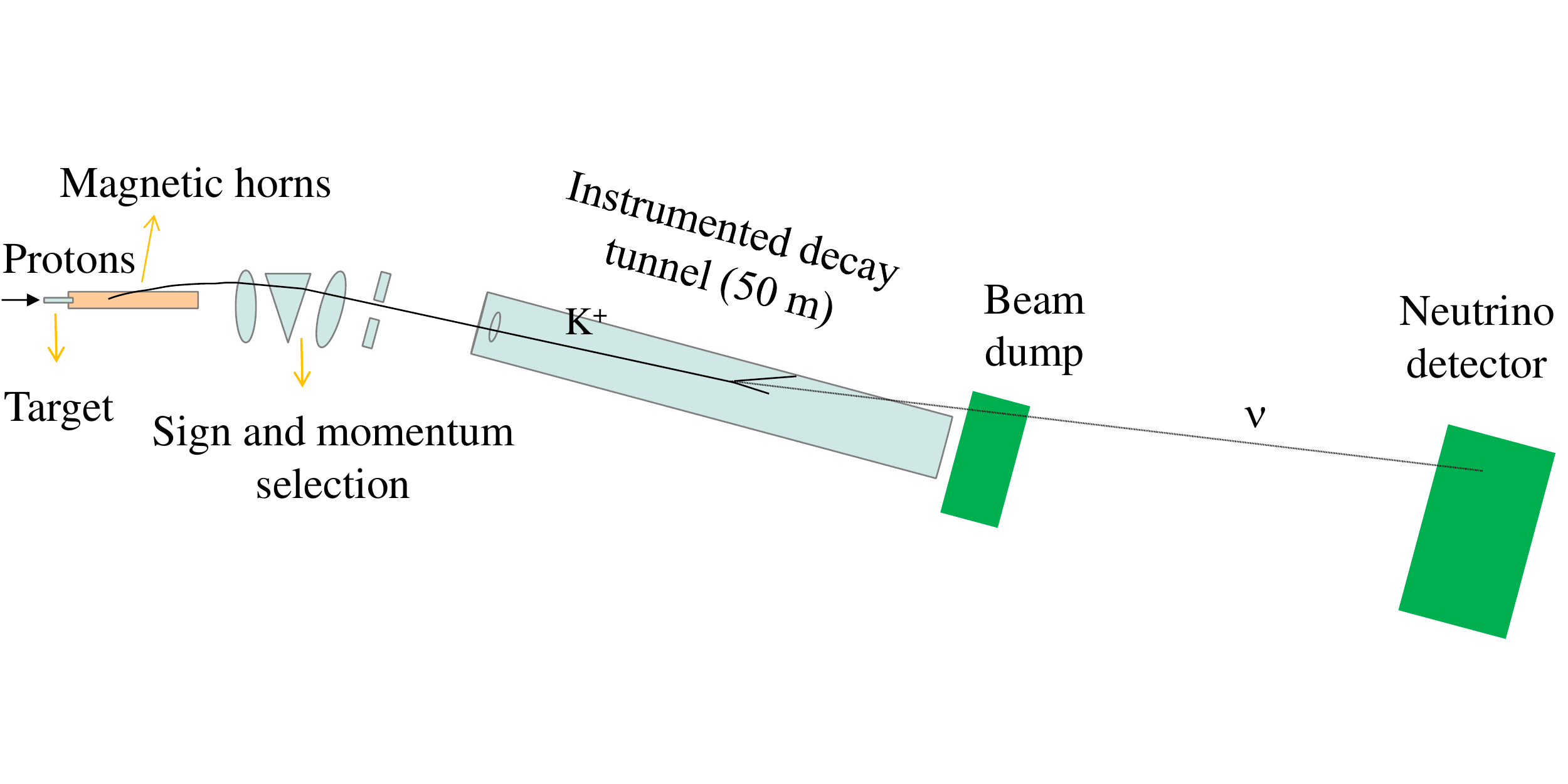} 
\caption{Layout of the facility (not to scale).}
\label{fig:schematics_2}
\end{figure}

The proposed facility is based on a conventional beam-line with 
primary protons impinging on a target, producing secondary hadrons which are
captured, sign selected and transported further down to the instrumented decay
tunnel (see Fig.~\ref{fig:schematics_2} and Sec.~\ref{sec:decaytunnel}). 
Inclusive secondary pion
yields on solid targets increase linearly with proton energy but the
technique presented in this study exploits high energy kaons to enhance
the $\pi^+/e^+$ separation at the calorimeters (Sec.~\ref{sec:backgr})
and to reduce the decay losses after the focusing system.  The optimal
value for the mean secondary momentum is around $8.5$~GeV.  Lower
values decrease the $\pi^+/e^+$ separation efficiencies, while higher
values reduce the flux and bring the $\nu_e$ spectrum above the region
of interest for future long-baseline experiments (0.5-4~GeV). 

In the following we assume to collect secondary positive particles
($\pi^+$, $K^+$) produced at the target and to transport them to the
entrance of the decay tunnel with a momentum bite of $\pm$20\%
centered at 8.5~GeV.  The decay tunnel consists of an evacuated
beampipe (40 cm radius) surrounded by the positron detectors (see
Sec.~\ref{sec:decaytunnel}). For the calculation of the neutrino flux,
we simulated pions and kaons distributed uniformly in a
$10\times10$~cm$^2$ window in the transverse plane and with a flat
polar angle distribution (up to 3~mrad). In fact, the actual meson
beam distribution at the entrance of the decay tunnel is not a
critical parameter because the neutrino beam divergence at this energy
is dominated by the large neutrino decay angle with respect to the
parent meson. The meson beam emittance has only to be small enough to
contain the secondary beam inside the tagging detector.  An unfocused
meson beam entering the decay tunnel within a window of $\pm$5~cm in
both transverse projections and with a polar angle smaller than
3~mrad, is fully contained in a 50~m long, 40~cm radius decay tunnel,
even including the tertiary muons produced in $\pi^+$ decays (see
Sec.~\ref{sec:decaytunnel}).  This phase space area corresponds to a
geometrical acceptance of the decay tunnel $A = 4 \cdot (5 \
\mathrm{cm}) \cdot (3 \ \mathrm{mrad}) = 4 \epsilon_{xx^{'}} = 4
\epsilon_{yy^{'}} = 0.60$~mm rad, where
$\epsilon_{xx^{'}}=\epsilon_{yy^{'}}=0.15$~mm rad in both transverse
projections.  Here $x$,$y$ and $x^{'}=dx/dz$, $y^{'}=dy/dz$ are
respectively the particle positions and slopes transverse to the
direction $z$ of the beam.

As discussed in Sec.~\ref{sec:double_tag}, a long extraction
($>$10~ms) is needed only for the event by event tag operation mode.
The capture of secondaries at the target in a facility operated in
event count mode can thus be implemented with conventional magnetic
horns.  On the other hand, fast extractions (10~$\mu$s) challenge the
positron tagger, whose local rate must be kept at the level of ${\cal
  O}(1)$~MHz/cm$^2$.  The optimal choice for the event count mode is
$\sim 2$~ms. Such extraction length has already been employed at the
CERN West Area Neutrino Facility (WANF~\cite{Heijne:1983rr}). It is
also the parameter on which the NUMI horns and their power supplies
have been originally designed~\cite{anderson}. Actually, in spite of
the fact that both NOVA and T2K implement a fast extraction
(10~$\mu$s) scheme~\cite{T2K_NBI}, the typical current pulse width
used to source their horns is $\sim2$~ms. Longer extractions introduce
additional constraints on the horn, due to the increase of Joule
heating. In particular, thicker conductors will be needed to reduce
resistive heating which, in turn, can cause beam deterioration due to
particle re-interactions~\cite{anderson}. In addition, long
extractions may reduce the $\nu_e$ CC purity due to cosmic background
at shallow depth (see Sec.~\ref{sec:rate}).

Downstream of the horn, the secondary beam is transported to the decay
tunnel entrance by a transfer line based on quadrupolar magnets for
the focusing and bending dipoles for the momentum selection.  Along
the transfer line, at the bending section, the high energy residual
primary protons are separated and transported to a dump~\cite{Adey:2013pio}.

In order to evaluate the secondary meson yields in this study we have
used Fluka 2011~\cite{Battistoni:2007zzb} to simulate primary proton
interactions
on a 110~cm long (about 2.6 interaction lengths) cylindrical beryllium
target of 3~mm diameter.  For the momentum bite considered, the
secondary yields at the target highly depend on the primary proton
energy.  We considered here proton energies of 30, 50, 60, 70, 120 and
450~GeV.  These correspond to facilities based on the JPARC proton
synchrotron (30~GeV), the upgrade of the U-70 accelerator in Protvino
(50-70~GeV)~\cite{ivanov}, the primary proton beamline of NUSTORM
(60~GeV)~\cite{Adey:2013pio}, the Main Injector at Fermilab (120~GeV)
used for the NUMI beam~\cite{numi}, the CERN-SPS operated in low
energy mode~\cite{Antonello:2012qx} (120~GeV) and the full energy
CERN-SPS~\cite{agarwalla::2013kaa} (450~GeV).

The capture and transfer line has not been simulated in this work as
it requires a site-dependent dedicated study that is beyond the scope
of the paper.  To evaluate the fluxes at the entrance of the decay
tunnel we used the phase space $xx^{'}$, $yy^{'}$ of pions and kaons
in a momentum bite of 8.5~GeV/c~$\pm$~20\% at 5~cm downstream the
110~cm long target.  We assume that all secondaries within an
emittance $\epsilon_{xx^{'}}=\epsilon_{yy^{'}}=0.15$~mm~rad are
focused with a typical horn focusing efficiency of
85\%~\cite{table_nustorm}. These particles are captured and
transported down to the entrance of the decay tunnel. The ellipse of
this area best matching the phase space distribution downstream the
target, i.e. the one maximising the pion flux, is selected and the
mesons lying within the ellipses in both transverse planes are
summed up.

The results are summarized in Table~\ref{tab:yield-horn}.  The second
and third columns show the pions and kaons per proton on target (PoT)
transported at the entrance of the decay tunnel. The fourth column
shows the number of PoT in a single extraction spill to obtain
$10^{10}$ pions per spill. The last column shows the number of
integrated proton on target that are needed to reach $10^4$ $\nu_e$ CC
events on a 500 tons neutrino detector (see Sec.~\ref{sec:rate}),
i.e. to enable a measurement of the $\nu_e$ absolute cross section
with a statistical precision of 1\%~\cite{note_stat}.  These proton fluxes are well
within the reach of the above-mentioned accelerators both in terms of
integrated PoT (from $5 \times 10^{20}$ at 30~GeV to $5 \times
10^{19}$ at 450~GeV) and protons per spill ($2.5 \times 10^{12}$ to $3
\times 10^{11}$). With respect to present running modes, two changes
have to be envisaged. The machine must provide proton pulses with ms
duration as in the former CERN-WANF (as already mentioned, current
neutrino beams are operated with pulse durations of $\sim
10$~$\mu$s). In addition, since the integrated number of spills is
large ($\sim 2 \times 10^8$ for proton pulses producing $10^{10}$
$\pi^+$ per spill - see third column of Table~\ref{tab:yield-horn}),
the accelerator should be run either with a repetition rate of several
Hz or in multi-turn extraction mode in order to have enough proton
bursts well separated in time hitting the target.

A notable exception is the U-70 synchrotron, which cannot be used in
its present form since the average power is less than 10~kW at 60~GeV
and the data taking (see Table~\ref{tab:yield-horn}) would exceed 6
years. The performance of a U-70 based facility will depend on the
final outcome of the OMEGA Project~\cite{ivanov}. An average power in
U-70 of 100~kW at 70 GeV would imply a $\sim$1 year long data taking
assuming an effective yearly run of 200 days at nominal power.  All
other accelerators considered in Tab.~\ref{tab:yield-horn} can be
employed without additional upgrades. Low energy drivers are, however,
slightly favored due to the higher repetition rate already available.

\begin{table}
\begin{center}
\begin{tabular}{ccccc}
\hline
\hline

$E_p$ (GeV) & $\pi^+$/PoT & $K^+$/PoT & PoT for a $10^{10}$ $\pi^+$  & PoT for $10^{4}$ $\nu_e$~CC \\
        &($10^{-3}$) &($10^{-3}$)&   spill  ($10^{12}$)           &  ($10^{20}$)        \\
\hline   
30      &	 4.0  &  0.39     &     2.5                  &    5.0              \\
50      &	 9.0  &  0.84     &     1.1                  &    2.4             \\
60      &	10.6  &  0.97     &     0.94                 &    2.0             \\
70      &	12.0  &  1.10     &     0.83                 &    1.76             \\
120     &	16.6  &  1.69     &     0.60                 &    1.16            \\ 
450     &	33.5  &  3.73     &     0.30                 &    0.52            \\
\hline   
\hline
\end{tabular}
\caption{Pion and kaon yields for horn focusing at
  (8.5$\pm$1.7)~GeV/c. The rightmost column is computed assuming a 500
  ton neutrino detector.}
\label{tab:yield-horn}
\end{center}
\end{table}

The magnetic horns cannot be pulsed for times much longer than 10~ms,
such as the long extraction needed to operate the tagged beam facility
in event by event mode.  An alternative to the horns for the mesons
capture is the use of purely static focusing and transport systems
based on large aperture quadrupoles/dipoles~\cite{Kopp:2006ky}.
In all these schemes, however, the capture is limited to the very
forward secondaries produced at target.  As a reference,
Table~\ref{tab:yield-forw} shows the pion and kaon yields within the
momentum bite, (8.5$\pm$1.7)~GeV/c and a forward 80~$\mu$Sr
acceptance~\cite{Ludovici:2010ci}.  Clearly, the large gain in flux
due to the horn-based focusing system compared with static systems
simplifies remarkably the design and construction of the event count mode
facility.

Since this angular acceptance is small, a Lithium lens could be
possibly used for the focusing of secondaries downstream the
target. However, operation of Li-lenses with the ${\cal
  O}(1)$~s extraction times needed for the event by event tag mode has
still to be demonstrated.

Compared to the yields of Table~\ref{tab:yield-horn} and
Table~\ref{tab:yield-forw}, we expect a reduction of the kaon yield
due to decay in the transport line ($\sim 16\% $ for an overall length
of 10~m) and finite capture and transport efficiency.  Similarly, the
use of graphite or INCONEL~\cite{Adey:2013pio} targets will increase
the secondary yield by 10-40\%. The yields of
Table~\ref{tab:yield-horn} thus represent an approximation of particle
production and transport down to the instrumented decay tunnel. The
precision is, however, appropriate for the aim of this study.

\begin{table}
\begin{center}
\begin{tabular}{ccccc}
\hline
\hline

$E_p$ (GeV) & $\pi^+$/PoT & $K^+$/PoT & PoT for a $10^{10}$ $\pi^+$ & PoT for $10^{4}$ $\nu_e$~CC \\
        &($10^{-3}$) &($10^{-3}$)&    spill ($10^{12}$)           &  ($10^{20}$)        \\
\hline   
30 &	0.24          & 0.027     & 42                       & 72                 \\
50 &	0.58          & 0.069     & 17                       & 28                 \\
60 &	0.73          & 0.091     & 14                       & 22                 \\
70 &	0.80          & 0.095     & 13                       & 20                 \\
120 &	1.25          & 0.16      & 8.0                      & 12.2                \\ 
450 &	3.65          & 0.43      & 2.7                      & 4.6                \\
\hline   
\hline
\end{tabular}
\caption{Pion and kaon yields forward (80~$\mu Sr$) at
  (8.5$\pm$1.7)~GeV/c. The rightmost column is computed assuming a 500
  ton neutrino detector.}
\label{tab:yield-forw}
\end{center}
\end{table}

\section{The instrumented decay tunnel}
\label{sec:decaytunnel}

\begin{figure}
\centering\includegraphics[width=\textwidth]{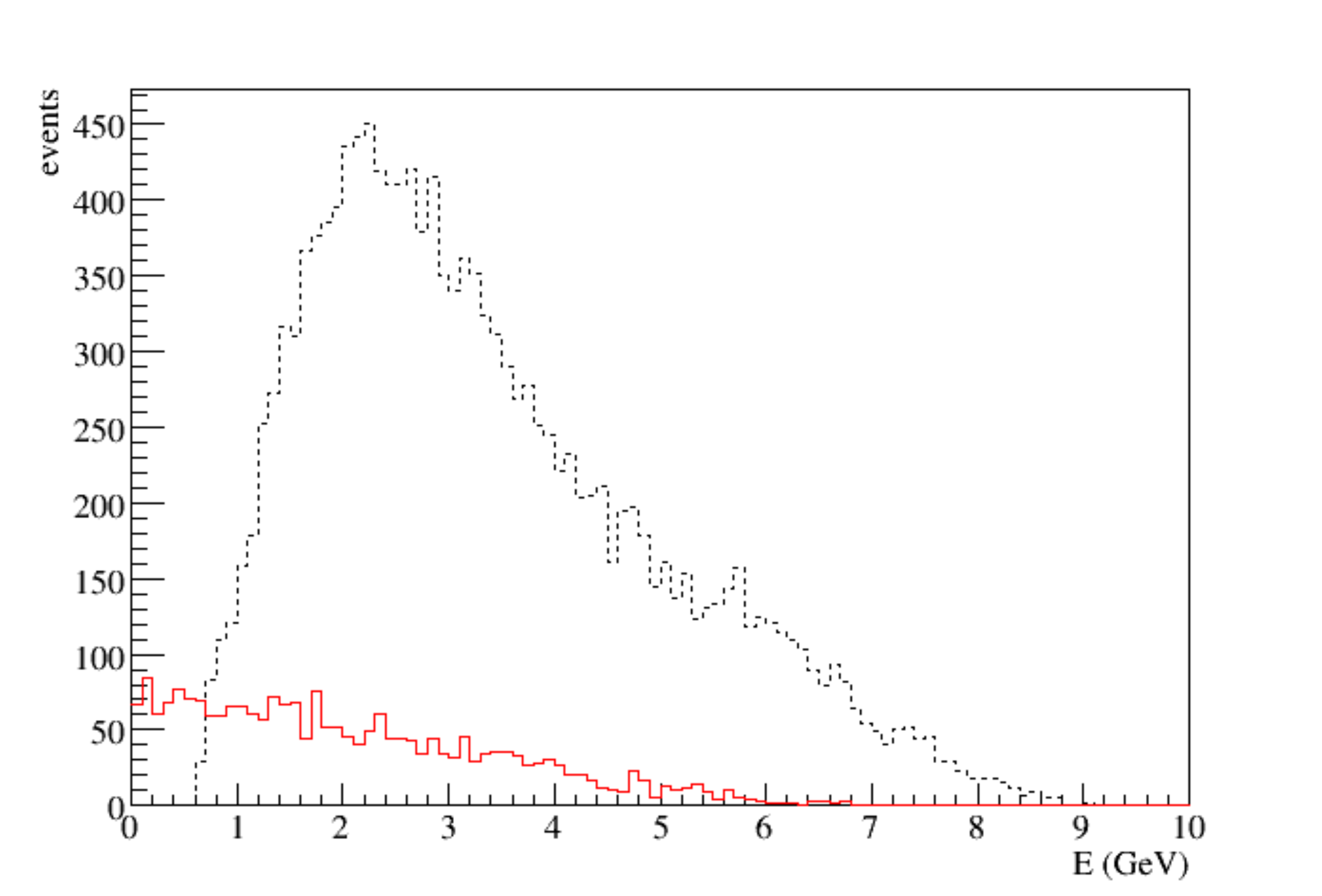} 
\caption{Energy distribution of positron (red continuous line) and
  pions (black dashed) from kaon decays hitting the calorimeter for
  $10^{5}$ $K^+$ at the entrance of the decay tunnel.}
\label{fig:energy_pos_pion}
\end{figure}

\begin{figure}
\centering\includegraphics[width=\textwidth]{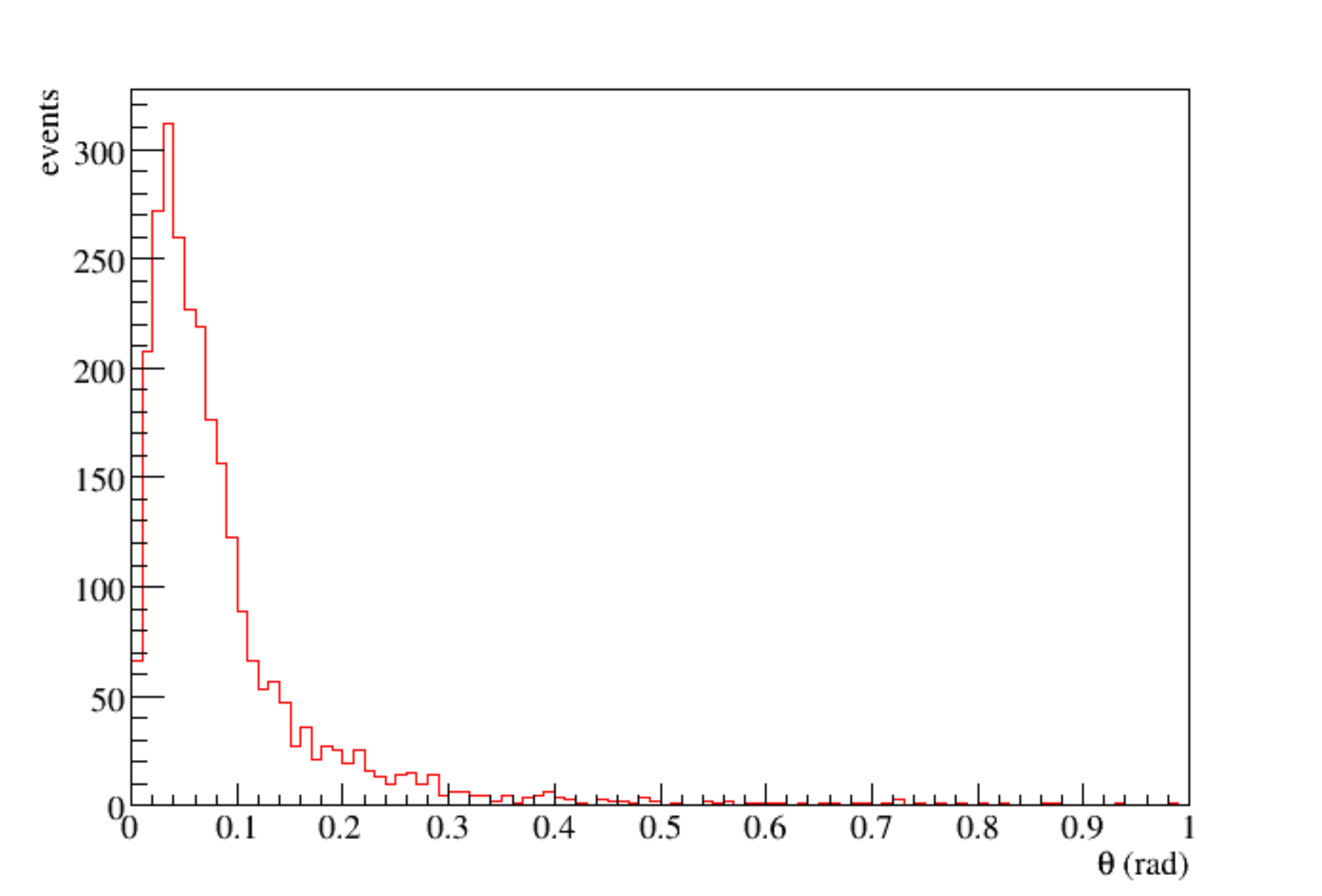} 
\caption{Polar angle distribution of positrons for $10^{5}$ $K^+$ at
  the entrance of the decay tunnel. Small angle positrons, i.e. $e^+$
  reaching the beam dump without crossing the calorimeter are
  included.}
\label{fig:theta_pos}
\end{figure}

\begin{figure}
\centering\includegraphics[width=\textwidth]{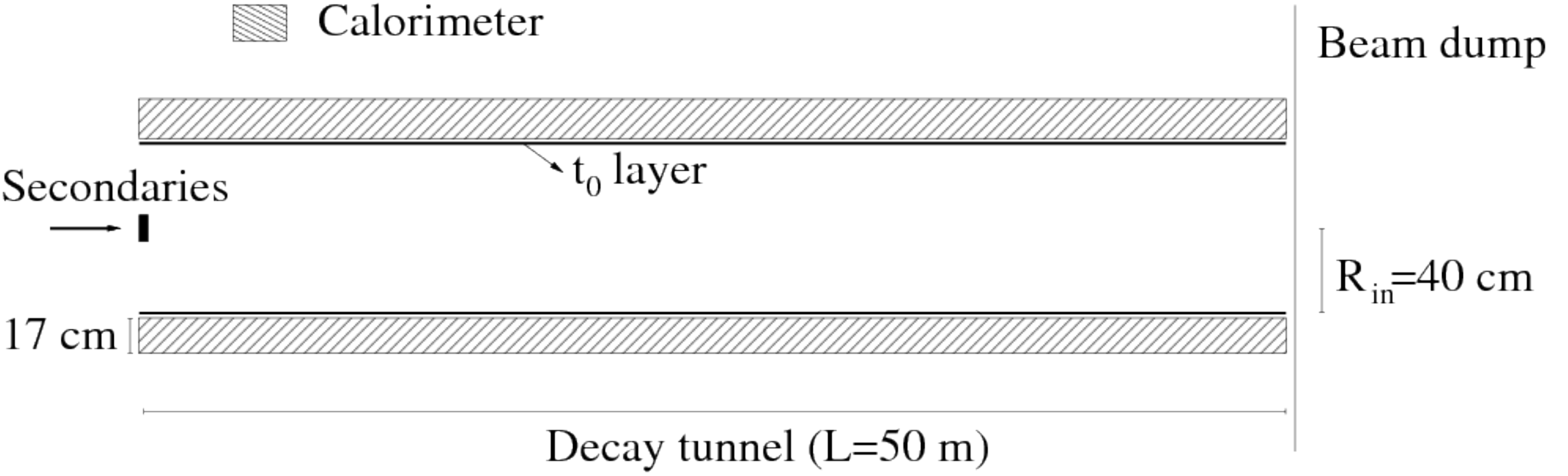} 
\caption{The instrumented decay tunnel (not to scale). The black
  rectangle on the left indicates the entrance window of the secondary
  particles in the transverse plane ($\pm$5~cm).}
\label{fig:scheme}
\end{figure}

The decay tunnel is a 50 m long evacuated beam pipe, surrounded by a
calorimeter that consists of a hollow cylinder with $R_{in}$ = 40~cm
inner radius and $R_{out}$ = 57~cm outer radius (see
Fig.~\ref{fig:scheme}).  The inner radius corresponds to a line of
sight between the entrance of the tunnel and the beam dump of 8~mrad.
All undecayed particles ($\pi^+$, $K^+$, $p$) and all muons from the
2-body decay of $\pi^+$ will reach the dump without crossing the
calorimeter. The overall rate at the calorimeter will therefore be
dominated by kaon decays. Since the decay products are forward going,
the calorimeter is thick enough to provide containment for
nearly all particles originating from kaon decays.  The energy and angle
distribution of the positrons from $K^+_{e3}$ decays is shown in
Figs.~\ref{fig:energy_pos_pion} and \ref{fig:theta_pos} (red
continuous line).  The mean polar angle of the positrons is 88~mrad.
Fig.~\ref{fig:energy_pos_pion} (black dashed line) also shows the
energy distribution of background $\pi^+$ from 2-body decay of $K^+$
(see Sec.~\ref{sec:backgr}).  Positrons in the decay tunnel are
identified by calorimetric techniques, exploiting the longitudinal
shower development for particle identification.  Photon rejection is
achieved by a ``$t_0$ layer'', a pre-shower that provides the absolute
time of arrival of the charged particle and is used to veto neutral
particles in the calorimeter.


\begin{figure}
\centering\includegraphics[width=\textwidth]{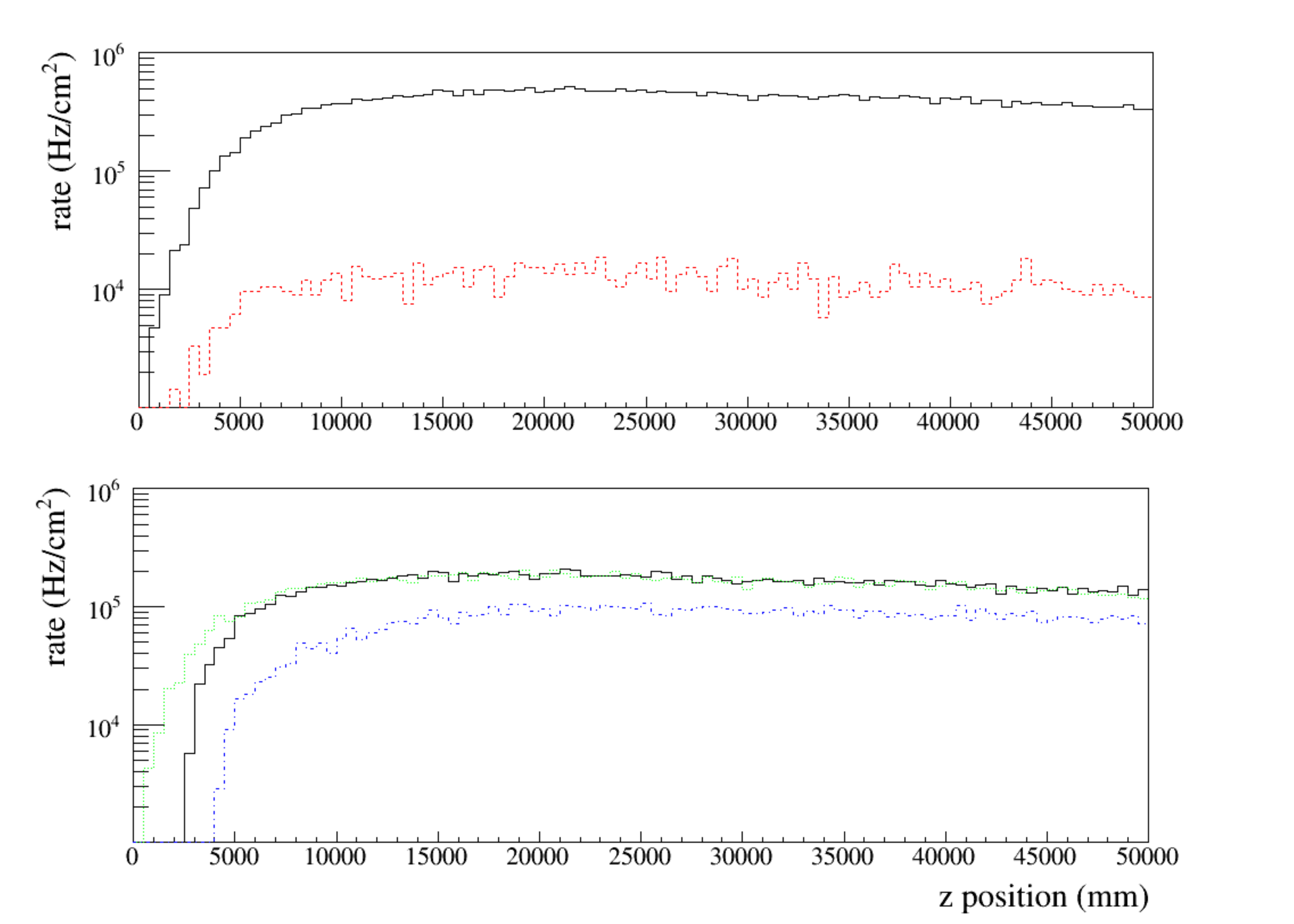} 
\caption{Upper plot. The black continuous (red dashed) line shows the
  overall particle (positron) rates in the calorimeter as a function
  of the $z$ position along the instrumented tunnel. Lower plot. Muon
  (black continuous), photon (green dotted) and pion (blue dot-dashed
  line) rates in the calorimeter as a function of the $z$
  position. Rates are computed for a 2~ms extraction length and
  $10^{10}$ $\pi^+$ per spill. }
\label{fig:rate_particles}
\end{figure}

Fig.~\ref{fig:rate_particles} shows the number of particles (expressed
in Hz/cm$^2$) entering the calorimeter as a function of the position
$z$ along the tunnel. Each bin corresponds to a surface of $2 \pi
R_{in} \Delta z = 1.26~\mathrm{m}^2$. For a 2~ms extraction length and
$10^{10}$ $\pi^+$ per spill, the maximum positron rate (upper plot -
red dashed line) is 10 kHz/cm$^2$. The overall rate (upper plot -
black continuous line) is dominated by muons originating by two-body
decays of kaons ($K_{\mu2} \equiv K^+ \rightarrow \mu^+ \nu_\mu$,
whose BR is $63.55 \pm 0.11\%$; see lower plot - black continuous
line) and photons (lower plot - green dotted line). The peak rate is
500~kHz/cm$^2$ (5~MHz per channel for a calorimeter with a granularity
of 10~cm$^2$). Due to the 3~mrad beam divergence and the Lorentz boost
of decayed particles, rates are low in the first 10~m of the tunnel
and they saturate at nearly constant value for $z>10$~m. During a
single 2~ms spill, we expect $10^{10}$ $\pi^+$ and $1.02 \times
10^{9}$ $K^+$ at the entrance of the tunnel (the $K^+/\pi^+$ ratio is
10.2\% for 120 GeV protons - see Tab.~\ref{tab:yield-horn}). The
number of kaon decays per spill is $5.6 \times 10^{8}$ (1 decay each
4~ps) and the corresponding number of positrons from $K_{e3}$ is $2.8
\times 10^{7}$.

The particle decays in the tunnel, the crossing of the $t_0$ layer and
the calorimeter response to charged and neutral particles have been
simulated through GEANT4~\cite{Agostinelli:2002hh,Allison:2006ve}.  In
this study the calorimeter is simulated as a homogenous copper
cylinder (radiation length $X_0=1.44$~cm, nuclear interaction length
$\lambda_I=15.3$~cm). Choices other than copper as the absorber
material (e.g. steel or lead-steel hybrid systems) are also worth
consideration in terms of cost-effectiveness, ease of machining and
nuclear properties.

In case of full longitudinal containment (electrons and pions), the
reconstructed energy $E_{tot}$ in the calorimeter is based on the true
particle energy smeared according to the following parametrization:
\be
\frac{\sigma_E}{E} = \frac{95\%} {\sqrt{E(\mathrm{GeV})}} \oplus 7\%  \ \ \mathrm{for \ hadrons}  
\ee
\be
\frac{\sigma_E}{E} = \frac{13\%} {\sqrt{E(\mathrm{GeV})}} \oplus 3\%  \ \ \mathrm{for} \ e^-,e^+,\gamma   
\ee
which correspond to typical performance of sampling calorimeters. Note
that the low-density active material is not simulated. Hence, the
actual outer radius of the calorimeter will be larger than $R_{out}$
(by $\sim 30\%$ for 1.5~cm Cu slabs interleaved by 0.5~cm scintillator
tiles).

The simulated energy deposition is sampled at the first $5X_0$ ($E_1$)
and $10X_0$ ($E_2$) and the variables $R_1 \equiv E_1/E_{tot}$ and
$R_2 \equiv E_2/E_{tot}$ are used for pion/positron separation. Since
uncertainties on $E_1$ and $E_2$ are completely dominated by
fluctuations due to lateral leakage, $E_1$ ($E_2$) is defined as the
energy deposited in the first 5 (10) radiation lengths inside a
cylinder of radius $2 R_M$ without additional smearing. $R_M$ =
1.568~cm is the Moliere radius of copper.  The material of the $t_0$
layer and beam pipe is neglected: the impact of the beam pipe material
on the background from photon convertion is discussed in
Sec.~\ref{sec:backgr}.

A positron is defined as an energy deposit in the calorimeter
associated with a hit in the $t_0$ layer. The energy deposited
$E_{tot}$ must be greater than $300$~MeV. We also request the energy
deposit in the first 5 and 10~$X_0$ to be significantly larger than
for a minimum ionizing particle (MIP): $R_1>0.2$, $R_2>0.7$.  These
requirements select positrons from $K_{e3}$ with 69\%
efficiency. Table~\ref{tab:poseff} summarizes the overall efficiency
(59\%) including the geometrical acceptance of the tagging calorimeter
due to positrons escaping at low polar angles into the beam dump.
Since local rates result only from kaons and are quite low compared
with collider requirements, several technologies are available both
for the calorimeter and for the $t_0$ layer. As a reference, we
considered a scintillator tile calorimeter readout by SiPM and WLS
fibers similar to the the CALICE AHCAL~\cite{adloff:2010hb} but with
much coarser longitudinal segmentation. Other options, developed both
for LHC and for CLIC are possible, too~\cite{cavallari}. Unlike
applications at colliders, pile-up mitigation and integrated doses are
not particularly critical. For a $R$ = 0.5~MHz/cm$^2$ local rate and a
tile size $S \simeq 10$~cm$^2$, the pile up probability is
\be 
P = R \ S \Delta T_{cal}
\ee
$\Delta T_{cal}$ being the recovery time of the calorimeter. It
corresponds to $P=0.05$ for $\Delta T_{cal}=10$~ns. In fact, pile-up
mostly results from the overlap of a muon from $K_{\mu2}$ with a
candidate positron.  Further pile-up mitigation is possible since
MIP-like deposits and punch-through particles can be vetoed or removed
offline using the longitudinal segmentation of the calorimeter and, if
needed, a muon catcher layer (Fe+muon chambers) located at
$R>R_{out}$.

Similarly, the integrated dose is not a critical parameter for this
facility.  From Table~\ref{tab:yield-horn}, 10$^4$ $\nu_e$ CC events
are obtained at the neutrino detector from the DIF of $1.94 \times
10^{17}$ kaons.  The deposited energy of decayed kaons is 150~MJ. Most
of this energy (64\%, i.e. the $K_{\mu2}$ BR) is either uniformely
distributed in the copper volume (muons) or lost outside the tunnel
(neutrinos). Assuming conservatively that all residual energy is
deposited in the first 3 $X_0$ of the calorimeter and that the
calorimeter extends from $z=10$ to $z=50$~m, the corresponding
integrated dose is $<1260$~Gray.

In event count mode, several technologies are available for the $t_0$
layer since the detector operates mainly as a photon tagger. For an
event by event tag facility (see Sec.~\ref{sec:double_tag}), however,
the $t_0$ layer must match or exceed the time resolution of the
neutrino detector. Plastic scintillator tiles offer $<1$~ns
resolution~\cite{Simon:2013zya} with ${\cal O}(10)$~cm$^2$
granularities. Conventional silicon detectors are not appropriate
because large surfaces increase the detector capacitance and
deteriorate the time resolution. Low gain avalanche
detectors~\cite{Cartiglia:2013haa} can overtake this limitation making
a semiconductor based $t_0$ layer a viable option.

In general, the technology choice will mostly be driven by cost
effectiveness. If the decay tunnel is instrumented from $z=10$ to
$z$=50~m, the corresponding calorimeter mass is 185~tons and the surface
of the $t_0$ layer is 100.5~m$^2$. For a 10~cm$^2$ granularity and
three longitudinal samples in the calorimeter, the overall
number of channels is thus $\sim 4 \times 10^{5}$.

\begin{table}
\begin{center}
\begin{tabular}{lc}
\hline
\hline
Cut & Efficiency  \\
\hline
$K_{e3}$ decay & 100\%  \\
$e^+$ in calorimeter & 85\%  \\
$R_1$,$R_2$ cuts & 67\% \\
$E_{tot}>300$~MeV & 59\% \\
\hline   
\hline
\end{tabular}
\caption{Positron efficiency after cuts.}
\label{tab:poseff}
\end{center}
\end{table}

\section{Background}
\label{sec:backgr}

The $K_{e3}$ branching ratio represents only $\sim$5\% of the overall
kaon decays. The bulk of particles crossing the calorimeter is due to
muons from the $K_{\mu2}$ decay mode and pions from the fully hadronic
mode ($K^+ \rightarrow \pi^+ \pi^0$ with $BR = (20.66 \pm
0.08)$\%~\cite{PDG}). The calorimetric muon/positron separation is
excellent since minimum ionizing  particles (muons and punch-through pions)
cluster at low values of $R_1$ and $R_2$. The misidentification rate
is below $10^{-3}$ when integrated to all muons produced by
$K_{\mu2}$, $K_{\mu3}$ and the DIF of pions from the other decay
modes. As a consequence, background from muon misidentification does
not represent a limitation for PID in the instrumented decay tunnel.

$\pi^+$/$e^+$ separation is much less efficient and dominates the
positron background. The main contribution comes from the two-body
fully hadronic decay mode $K^+ \rightarrow \pi^+ \pi^0$. The spectrum
of charged pions from this decay mode is shown in
Fig.~\ref{fig:energy_pos_pion} together with the positron
signal. Again, pions cluster at lower values of $R_{1,2}$ but charge
exchange and the intrinsic fluctuation of the e.m. component in
hadronic showers can mimic a positron, especially at low energy. The
integrated misidentification probability is $\epsilon_{\pi^+
  \rightarrow e^+} = 2.2$\% and the contamination in the positron
signal due to pion misidentification amounts to 13\%. Possible
overlaps between the photon from $\pi^0$ decay and the pions are
included in the simulation and, as well as Dalitz $\pi^0$ decays, 
give negligible contributions. Pile-up has not been included since
its effect is marginal (see Sec.~\ref{sec:decaytunnel}).

Additional contributions to background come from the $K^+ \rightarrow
\pi^+ \pi^+ \pi^- $ decay mode (BR~$\simeq$~5.6\%). In spite of the
smaller BR, the higher charge multiplicity results into a larger
misidentification efficiency and increases the overall background
contamination to 18\%.  Unlike $K^+ \rightarrow \pi^+ \pi^0$, the $K^+
\rightarrow \pi^+ \pi^+ \pi^- $ three-prong decays can be vetoed
requiring no identified charged pions pointing to the decay
vertex. This background reduction technique, however, needs to exploit
the granularity of the detectors to identify the decay vertex along
$z$ and the time resolution of the $t_0$ layer. For the case under
study (2~ms extraction and 10~cm$^2$ granularity), the precision on
$z$ is $\sim$1~m and the average time among $K^+$ decays into
charged pions in a 1~m section of the tunnel is 510~ps. As a
consequence, multi-prong background reduction sets the scale of the
$t_0$ layer time resolution to $\mathcal{O}(100)$~ps. In the present
analysis, this reduction technique has not been considered.

Photons produced in the decay tunnel originate from $\pi^0$ decays in
the semileptonic and hadronic modes of the kaons. The most important
source is $K^+ \rightarrow \pi^+ \pi^0$. These photons will not give a
hit in the $t_0$ layer associated with the energy deposit in the
calorimeter. 
Photons converted in the material inside the $t_0$ layer can constitute a
background because the event count mode facility does not exploit time
correlation among particles  ($\pi^+$ and $\gamma$'s in this case) or
the other particles can lay outside the geometrical acceptance of the
calorimeter. For a 1.5~mm Be beam-pipe~\cite{ATLAS:2010aba}, the
conversion rate is $3 \times 10^{-3}$ and, even without additional
background mitigation techniques, the contamination is less than
2\%. It grows to 6\% for a 1~mm Al vacuum tanks. Again, if the $t_0$
layer has a time resolution of $\mathcal{O}(100)$~ps, this background
can be suppressed to a negligible level vetoing prompt pions that
originate from the same area of the candidate positron. The photon
background is also negligible if the $t_0$ layer is installed inside
the vacuum pipe, as for the Large Angle Veto calorimeters of
NA62~\cite{NA62_TDR}. All sources of background are summarized in
Tab.~\ref{tab:background} together with the $\epsilon_{\pi^+
  \rightarrow e^+}$ misidentification probability.

\begin{table}
\centering
\begin{tabular} { c c c c c}
\hline
 Source  & BR   & Misid & $\epsilon_{X \rightarrow e^+}$ & Contamination \\
\hline
$\pi^+ \rightarrow \mu^+ \nu_\mu$ & 100\% & $\mu \rightarrow e$ misid. & $<$0.1\%  & neglig. (outside acceptance) \\
$\mu^+ \rightarrow e^+ {\bar \nu_\mu} \nu_\mu$ & DIF & genuine $e^+$ & $<$0.1\% & neglig. (outside acceptance) \\
$K^+ \rightarrow  \mu^+ \nu_\mu$  & 63.5\% & $\mu \rightarrow e$ misid. & $<$0.1\% & negligible \\
$K^+ \rightarrow  \pi^+ \pi^0$  & 20.7\% & $\pi \rightarrow e$ misid. & 2.2\% & 13\% \\
$K^+ \rightarrow  \pi^+ \pi^+ \pi^-$  & 5.6\% & $\pi \rightarrow e$ misid. & 3.8\% & 5\% \\
$K^+ \rightarrow  \pi^0 \mu^+ \nu_\mu$  & 3.3\% & $\mu \rightarrow e$ misid. & $<$0.1\% & negligible \\
$K^+ \rightarrow  \pi^+ \pi^0 \pi^0$  & 1.7\% & $\pi \rightarrow e$ misid. & 0.5\% & negligible \\
\hline
\end{tabular}
\label{tab:background}
\caption{Sources of background and misidentification probability.}
\end{table}

\section{Rates at the neutrino detector and systematic errors }
\label{sec:rate}

The beamline and instrumented tunnel of Secs.~\ref{sec:beamline} and
\ref{sec:decaytunnel} produce a neutrino beam that is enriched in
$\nu_e$ from kaon decays and depleted in $\nu_e$ from muon
DIF. Assuming a far detector located 100~m from the entrance of the
tunnel (50~m from the beam dump), 500~ton mass (isoscalar target) and
a cross-sectional area\footnote{It corresponds to the surface of a
  cylindrical detector with a geometrical acceptance of 100~mrad.}  of
17.7$\times$17.7~m$^2$, $2.1 \times 10^{-3}$~$\nu_\mu$/PoT
cross the detector for 120~GeV protons. The number of electron
neutrinos from $K^+$ decays is $3.8 \times 10^{-5}$~$\nu_e$/PoT and the
$\nu_e$ from DIF is $1.2 \times 10^{-6}$~$\nu_e$/PoT. Since
the $K^+/\pi^+$ ratio is nearly constant from 30 to 450~GeV (see
Tab.~\ref{tab:yield-horn}), the $\nu_e/\nu_\mu$ flux ratio at the neutrino
detector is independent of the proton energy and is:
$$
\frac{\Phi_{\nu_e}}{\Phi_{\nu_\mu}} = 1.8 \ \% \ (\nu_e \ \mathrm{from} \ K_{e3}) 
\ \ ; \ \  
\frac{\Phi_{\nu_e}}{\Phi_{\nu_\mu}} = 0.06 \ \% \ (\nu_e \ \mathrm{from \ DIF}) 
$$
As expected, the beam is enriched in $\nu_e$ from kaon decays, while
the contamination of $\nu_e$ from DIF is negligible. The
positrons at the calorimeter are therefore proportional to the number
of electron neutrino crossing the detector. The small difference from the
na\"ive scaling of Fig.~\ref{fig:scaling} is mostly due to $\nu_\mu$ from $K^+$
decays (not included in Eq.~\ref{eq:scaling}).

\begin{figure}
\centering\includegraphics[width=\textwidth]{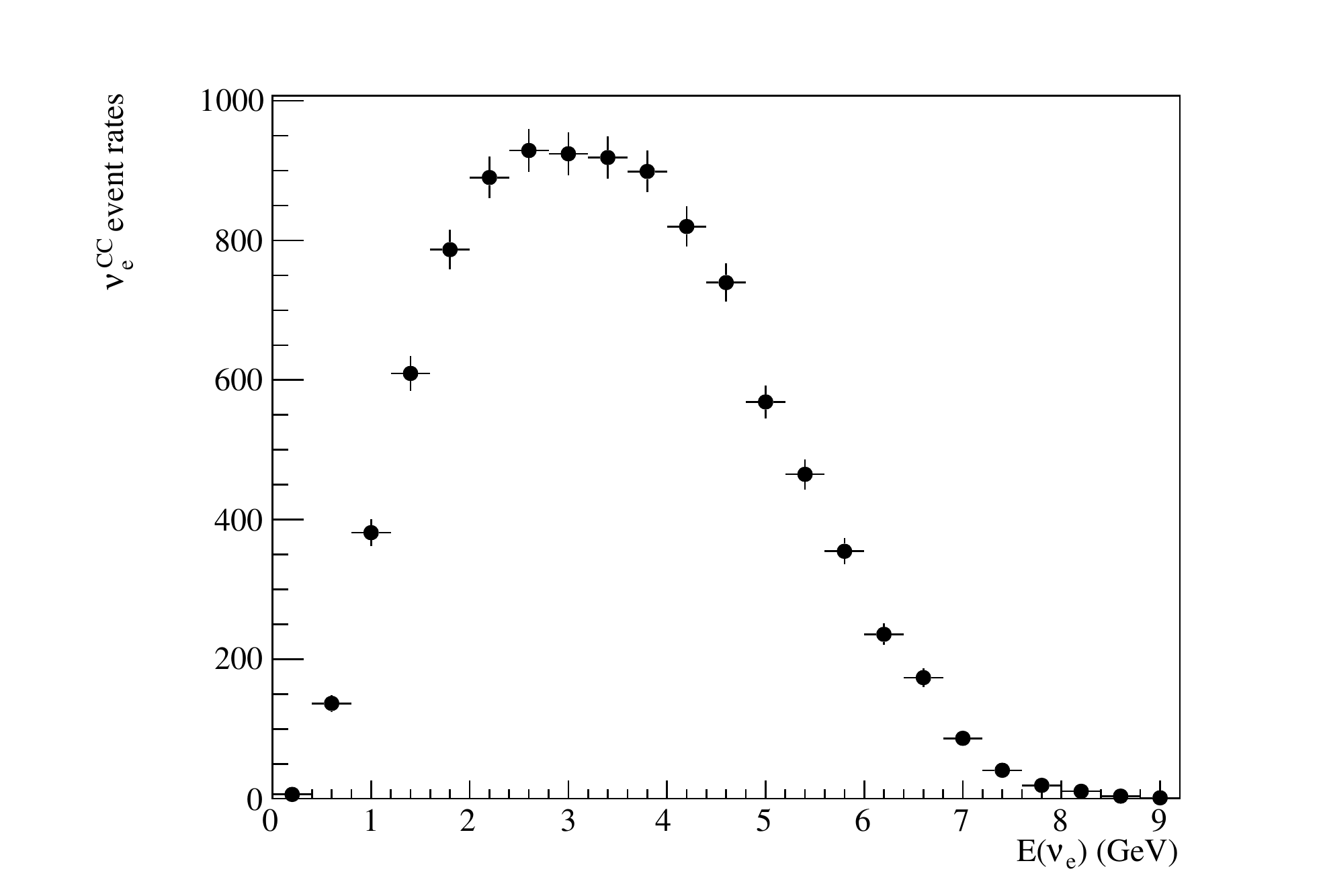} 
\caption{Energy distribution of the $\nu_e$~CC events.}
\label{fig:nue_rates}
\end{figure}

The event rate has been estimated folding the incoming flux with the
corresponding $\nu_e$~CC cross-section. The spectrum of $\nu_e$~CC
events at the detector for $E_\nu > 0.3$~GeV is shown in
Fig.~\ref{fig:nue_rates} (the events in the first bin correspond to
$0.3 < E_\nu< 0.4$~GeV). The mean energy is 3~GeV with a FWHM of
$\sim$3.5~GeV.

The number of positrons reconstructed in the calorimeter is directly
proportional to the flux of $\nu_e$ at the source.  This provides a direct
measurement of the $\nu_e$ flux, independent of the hadron
production yield, the $K/\pi$ ratio, the secondary transport
efficiency and the number of integrated PoT, i.e. of the main source
of flux systematic errors in cross section measurements. It depends,
however, on the geometrical efficiency of the neutrino detector, the
knowledge of the positron efficiency in the calorimeter and on the
background. In fact, only $\sim$80\% of the tagged positrons will
produce neutrinos that cross the detector and $\sim$15\% of the
$\nu_e$~CC observed at the detector will remain untagged since the
corresponding positron is lost in the beam dump. In turn, this implies
that the geometrical efficiency will also depend on the kinematics of
$K_{e3}$ decay and on the actual divergence of the beam at the
entrance of the tunnel.  Finally, the geometrical efficiency slightly depends
on the slope of the hadron energy distribution in the momentum
bite. 

All beam parameters describing the spatial distribution of kaons at
the entrance of the tunnel can be measured monitoring the charged
pions in dedicated low-intensity proton extractions with negligible
statistical uncertainty. As for standard collider applications, the
PID separation capability of the calorimeter will be measured in
test-beams before the installation and can be cross-checked
on site. Although a detailed assessment of systematics requires a full
simulation of the beamline and the detector response, to best of
current knowledge the overall systematic budget can be kept within
${\cal O}(1\%)$.

If the facility is operated in event count mode, the time resolution
constraint on the neutrino detector are loose and the technology
choice is mostly driven by the neutrino detection efficiency and the
corresponding systematics~\cite{Abe:2007bi}. For electron
identification, scintillator based detectors offer fast time response
($<$10~ns) and good energy resolution but the granularity and PID
capability is limited by the size of the scintillator cells.

Liquid Argon (LAr) detectors have superior granularity and PID capabilities, 
thus achieving a smaller systematic error associated to $\pi^0$ 
mis-identification background. On the other hand, the longer integration time 
in LAr detectors results in pile-up of signal events and cosmic background. 
In particular the proton extraction length (2~ms) matches the integration time
of LAr detectors. Unlike fast extraction beams, the timing of the event with the
proton current profile will be less effective for cosmic rejection even
if a scintillator based fast trigger~\cite{Antonello:2014rwa} is used.
According to~\cite{Rubbia:2014dva}, LAr detectors equipped with an active
veto system can be operated even with moderate overburden and
sub-GeV electron neutrinos. 

Finally, it is worth mentioning that a facility operated in event
count mode can be run with reversed polarity to measure
the $\bar{\nu}_e$ cross section. In addition, the high energy
$\nu_\mu$~CC subsample is mostly due to the $K_{\mu 2}$ decays; it can
thus be employed in combination with the positron rate (or, more
directly, with the tagged large-angle muon rate) to retrieve
information on the $\nu_\mu$ CC cross section. These applications have
not been considered in the present study.

\section{Event by event tag}
\label{sec:double_tag}

The beam parameters of Sec.~\ref{sec:beamline} cannot be used to run
the facility in event by event tag mode. For a 2~ms proton extraction,
one decay every 4~ps will be observed on average at the tagging
detector. The average time difference between positrons would be
$\sim 70$~ps, which corresponds to a time resolution at the limit both
of current technologies and of the intrinsic limitation of this
method (see below).

In the event by event tagging facility, the time coincidence is
performed between the timing of the neutrino interaction and the
timing of the positron.  Since the neutrino production vertex is
unknown, the timing difference is corrected for the time of flight
between the neutrino interaction vertex and the position of the
positron tag in the decay tunnel.  A neutrino is uniquely associated
to a positron, i.e. it is flavor tagged as an electron neutrino on an
event by event basis, if the time difference $\delta t$ between the
tagging detector and the neutrino detector is compatible with
$\Delta/c$ within the timing uncertainties.  Here, $\Delta$ is the
distance between the neutrino interaction vertex and the point along
the decay tunnel axis at the same longitudinal position as the
positron impact point on the tagging detector.  Since both the
neutrino and positron emission angles are small, this is a good
approximation of the neutrino position at the time of the positron
tagging.  For a tagging calorimeter of inner radius $R_{in}$, the
average correction to $\Delta$ due to the positron emission angle
would be of the order of $O(R_{in}\theta/2 c)$~ps, or $\sim 80$~ps for
an average positron angle of $\theta=88$~mrad.  The intrinsic limit of
the time coincidence can be conservatively assumed of the same order,
due to the spread of the positron emission angle distribution.  The
uncertainty in the time coincidence due to detector resolution of the
neutrino vertex and positron tagging positions can be estimated in
$O(50)$~ps.

The requirements on the timing resolution can be loosened well above
the intrinsic limit increasing the proton extraction length. A 1~s
extraction would bring the average time between two decays to 1.5~ns
and the average time among $K_{e3}$ decays to 30~ns.  An event by
event tag facility could hence be designed employing existing
technologies. Defining $\delta$ as the linear sum of the $t_0$-layer
and neutrino detector time resolution, the accidental tag probability
$\mathcal{A}$ is
\be 
\mathcal{A} \equiv \left[ N_K \cdot \mathrm{BR} (K_{e3}) (1- e^-\frac{\gamma_K c
\tau_K}{L}) \epsilon + \mathrm{bkg} \right] \cdot \delta \simeq 2 \times 10^{7} \frac{\delta}{T_{extr}}
\label{eq:accidental}
\ee
where $N_K$ the number of kaons per second, $L$ the length of the
decay tunnel, $\epsilon$ the overall tagging efficiency, ``bkg'' the
background contamination and $T_{extr}$ the proton extraction length.
Eq.~\ref{eq:accidental} sets the scale for the overall time resolution
that is needed to build a tagged neutrino beam with rates at the
detector similar to the event count facility discussed
above. $\mathcal{A}$ is 2\% for $\delta \simeq$1~ns.

An event by event tag facility offers several advantages, including
the possibility to veto the intrinsic contamination of conventional
beams for every observed event at the far
detector~\cite{Ludovici:1996sx} and to measure the neutrino energy
reconstructing the kinematics of $K_{e3}$. It can also be used to
measure the $\nu_\mu$ cross section from the $K^+ \rightarrow \mu^+
\nu_\mu$ decays, tagging the large angle muons in the decay tunnel
and counting the $\nu_\mu$~CC events in the detector occurring at $t
\simeq \Delta/c$. For the beam parameters considered above, 70\% of the
events can be fully reconstructed observing the $\gamma$ pair in
coincidence with the positron. For these events, the neutrino energy
resolution $\sigma_{E_\nu}$ is $\sigma_{em} \oplus
\Delta_p/\sqrt{12}$, where $\sigma_{em}$ is the e.m. energy resolution
of the calorimeter and $\Delta_p$ is the momentum bite of the
beamline. For the beam parameters of Sec.~\ref{sec:beamline}, this
accuracy is dominated by $\Delta_p$ and $\sigma_{E_\nu} = ( 0.35
\oplus 0.49)\ \mathrm{GeV} = 0.6$~GeV for 3~GeV neutrinos. I.e. a
momentum bite of 15\% limits the relative precision of the neutrino
energy reconstruction to $\sim 15$\%.

An event by event tag facility must meet several challenges. An
increase of the extraction time up to $T_{extr} \simeq 1$~s makes the
use of conventional horns unpractical. The focusing system will hence
rely on static components, as discussed in Sec.~\ref{sec:beamline}.
In addition, the momentum bite must be significantly smaller than the
event count facility to fully exploit the $K_{e3}$ kinematic
reconstruction.  Finally, due to the substantial increase in
extraction time (1~s versus 2~ms), the cosmic ray background in the
neutrino detector is $\mathcal{O}(10 \times )$ the background in
event count  mode.  Since a $\nu_e$ candidate must match a positron
candidate at the calorimeter within $\delta$, the cosmic background
contamination scales as $\mathcal{A} \cdot T_{2}/T_{1}$. Here, $T_1$
and $T_2$ are the extraction times in event count  and event by event
tag mode, respectively. For $T_1=2$~ms, $T_2=1$~s and
$\mathcal{A}=0.02$ the cosmic background increases by $\sim 10$.

In general, the event by event tag facility poses stronger technical
challenges and, unlike the event count mode, its design will require a
significant R\&D phase.
 
\section{Conclusions}
\label{sec:conclusions}

Three body $K^+ \rightarrow e^+ \pi^0 \nu_e$ decays in conventional
neutrino beams offer unique opportunities to measure the $\nu_e$
charged current cross section with a precision of $\sim 1\%$. In this
paper, we discussed a facility that identifies positrons in the decay
tunnel using calorimetric techniques to tag the production of $\nu_e$
at source. The positron rate at the instrumented decay tunnel removes
the most important systematics related with the knowledge of the
initial flux. An overall tagging efficiency of 59\% is achievable in a
specific beamline configuration that enhances the $\nu_e$/$\nu_\mu$
ratio to $\sim 2\%$ and reduces the $\nu_e$ contribution due to DIF to
$<0.1\%$.  Local rates and pile-up are well below the critical values
for conventional calorimeters working at colliders. The integrated
dose corresponding to $10^4$ events observed at the neutrino detector
does not exceed 1.3~kGy. The construction of this facility, which
monitors the positron production but does not associate uniquely the
positron to the observed $\nu_e$ (``event count mode''), can be
accomplished using existing technologies. For a 0.5~kton neutrino
detector, the beam intensity needed to reach the 1\% precision is well
within reach of proton accelerators at CERN, Fermilab and JPARC.

The corresponding setup operated in event by event tag mode has been
discussed, too.  Event by event tagging, however, requires a purely
static focusing system, a reduction on the secondary momentum bite and
an overall time resolution $\delta$ of ${\cal O}(1)$~ns. Its
implementation therefore implies additional R\&D and advances in beam
and detector technologies.

\section*{Acknowledgments}
The authors gratefully acknowledge discussions and suggestions from
K.~Anderson, A.~Bross, N.~Cartiglia, M.~Dracos, J.~Hylen, P.~Loverre,
M.~Mezzetto, J.~Morfin, B.~Popov, C.~Rubbia and T.~Tabarelli.



\begin{thebibliography}{999}

\bibitem{Formaggio:2013kya}
  J.~A.~Formaggio and G.~P.~Zeller,
  Rev.\ Mod.\ Phys.\  {\bf 84} (2012) 1307.

\bibitem{Alvarez-Ruso:2014bla}
  L.~Alvarez-Ruso, Y.~Hayato and J.~Nieves,
  New J.\ Phys.\  {\bf 16} (2014) 075015.

\bibitem{Gran:2006jn}
  R.~Gran {\it et al.}  [K2K Collaboration],
  Phys.\ Rev.\ D {\bf 74} (2006) 052002.   R.~Gran  {\it et al.} [K2K Collaboration],
  Nucl.\ Phys.\ Proc.\ Suppl.\  {\bf 221} (2011) 98.

\bibitem{Adamson:2009ju}
  P.~Adamson {\it et al.}  [MINOS Collaboration],
  Phys.\ Rev.\ D {\bf 81} (2010) 072002.

\bibitem{Dobson:2013uxa}
  J.~Dobson [T2K Collaboration],
  Nucl.\ Phys.\ Proc.\ Suppl.\  {\bf 237-238} (2013) 199.

\bibitem{Abe:2013jth}
  K.~Abe {\it et al.}  [T2K Collaboration],
  Phys.\ Rev.\ D {\bf 87} (2013) 092003.


\bibitem{Abe:2014nox}
  K.~Abe {\it et al.}  [T2K Collaboration],
  Phys.\ Rev.\ D {\bf 90} (2014) 052010.

\bibitem{Nakajima:2010fp}
  Y.~Nakajima {\it et al.}  [SciBooNE Collaboration],
  Phys.\ Rev.\ D {\bf 83} (2011) 012005.

\bibitem{Tice:2014pgu}
  B.~G.~Tice {\it et al.}  [MINERvA Collaboration],
  Phys.\ Rev.\ Lett.\  {\bf 112} (2014) 231801.

\bibitem{Acciarri:2014isz}
  R.~Acciarri {\it et al.}  [ArgoNeuT Collaboration],
  Phys.\ Rev.\ D {\bf 89} (2014) 112003.

\bibitem{review_xsect} For a review of latest results see 
F. Sanchez, Talk at XXVI International Conference on Neutrino Physics and Astrophysics, June 2-7, 2014, Boston, MA, US.

\bibitem{Blietschau:1977mu}
  J.~Blietschau {\it et al.}  [Gargamelle Collaboration],
  Nucl.\ Phys.\ B {\bf 133} (1978) 205.

\bibitem{Abe:2014agb}
  K.~Abe {\it et al.}  [T2K Collaboration],
  Phys.\ Rev.\ Lett.\  {\bf 113} (2014) 241803.

\bibitem{Day:2012gb}
  M.~Day and K.~S.~McFarland,
  Phys.\ Rev.\ D {\bf 86} (2012) 053003.

\bibitem{Dusini:2012vc}
  S.~Dusini, A.~Longhin, M.~Mezzetto, L.~Patrizii, M.~Sioli, G.~Sirri and F.~Terranova,
  Eur.\ Phys.\ J.\ C {\bf 73} (2013)  2392.

\bibitem{Coloma:2012ji}
  P.~Coloma, P.~Huber, J.~Kopp and W.~Winter,
  Phys.\ Rev.\ D {\bf 87} (2013)  033004.

\bibitem{Volpe:2003fi}
  C.~Volpe,
  J.\ Phys.\ G {\bf 30} (2004) L1.

\bibitem{McLaughlin:2004va}
  G.~C.~McLaughlin,
  Phys.\ Rev.\ C {\bf 70} (2004) 045804.

\bibitem{Oldeman:2009wa}
  R.~G.~C.~Oldeman, M.~Meloni and B.~Saitta,
  Eur.\ Phys.\ J.\ C {\bf 65} (2010) 81.


\bibitem{Adey:2013pio}
  D.~Adey {\it et al.}  [nuSTORM Collaboration],
  arXiv:1308.6822 [physics.acc-ph].

\bibitem{Spitz:2014hwa}
  J.~Spitz,
  Phys.\ Rev.\ D {\bf 89} (2014) 073007.

\bibitem{hand1969}
L. N. Hand, ``A study of 40-90 GeV neutrino interactions using a tagged neutrino
beam,'' Proceedings of Second NAL Summer Study, Aspen, Colorado, 9 Jun - 3
Aug 1969, p.37.

\bibitem{Pontecorvo:1979zh}
  B.~Pontecorvo,
  Lett.\ Nuovo Cim.\  {\bf 25} (1979) 257.

\bibitem{denisov}
P. Denisov et al., preprint IHEP 81-98, Serpukhov, 1981.

\bibitem{bernstein}
R.H.~Bernstein et al., FERMILAB-Proposal-0788, 1989.

\bibitem{Ludovici:1996sx}
  L.~Ludovici and P.~Zucchelli,
  [hep-ex/9701007].

\bibitem{Ludovici:2010ci}
  L.~Ludovici and F.~Terranova,
  Eur.\ Phys.\ J.\ C {\bf 69} (2010) 331.

\bibitem{Geer:1997iz}
  S.~Geer,
  Phys.\ Rev.\ D {\bf 57} (1998) 6989
   [Erratum-ibid.\ D {\bf 59} (1999) 039903].

\bibitem{Zucchelli:2002sa}
  P.~Zucchelli,
  Phys.\ Lett.\ B {\bf 532} (2002) 166.

\bibitem{Baldy:1999dc}
  R.~Baldy, J.~L.~Baldy, A.~E.~Ball, P.~Bonnal, M.~Buhler-Broglin, C.~Detraz, K.~Elsener and A.~Ereditato {\it et al.},
  CERN-SL-99-034-DI, CERN-SL-99-34-DI, INFN-AE-99-05, INFN-AE-99-5.

\bibitem{PDG}
 K.~A.~Olive {\it et al.}  [Particle Data Group Collaboration],
  Chin.\ Phys.\ C {\bf 38} (2014) 090001.

\bibitem{Heijne:1983rr}
  E.~H.~M.~Heijne,
  CERN-83-06, CERN-YELLOW-83-06.

\bibitem{anderson} K.~Anderson and J.~Hylen, personal
  communication. See also K.~Anderson, ``NuMI/NOvA 700kW Horn 1
  Stripline Vibration Measurements'', Talk at 9$^{th}$ International
  Workshop on Neutrino Beams and Instrumentation (NBI 2014).

\bibitem{T2K_NBI}
T.~Sekiguchi, ``T2K Horn Status'',  Talk at 9$^{th}$ International
Workshop on Neutrino Beams and Instrumentation (NBI 2014).  

\bibitem{Battistoni:2007zzb}
  G.~Battistoni, S.~Muraro, P.~R.~Sala, F.~Cerutti, A.~Ferrari, S.~Roesler, A.~Fasso and J.~Ranft,
  AIP Conf.\ Proc.\  {\bf 896} (2007) 31; A.~Ferrari, P.~R.~Sala, A.~Fasso and J.~Ranft,
  CERN-2005-010, SLAC-R-773, INFN-TC-05-11. Available at \texttt{http://www.fluka.org} 

\bibitem{ivanov} 
S.~Ivanov, ``Accelerator Complex U70 of
IHEP-Protvino: Status and Prospects for Upgrade'', Talk at 16$^{th}$
Lomonosov Conference on Elementary Particle Physics, August 22-28
2013, Moscow, RU.

\bibitem{numi}
K. Anderson {\it et al.}, FERMILAB-DESIGN-1998-01 (1998).

\bibitem{Antonello:2012qx}
A. Antonello {\it et al.},
arXiv:1208.0862 [hep-ph].

\bibitem{agarwalla::2013kaa} 
  S.~K.~Agarwalla {\it et al.}  [LAGUNA-LBNO Collaboration],
  JHEP {\bf 1405} (2014) 094.

\bibitem{table_nustorm} See e.g. Tables VIII-X of Ref.~\cite{Adey:2013pio}.

\bibitem{note_stat} Larger statistics may be needed to measure
  differential cross sections with a single bin precision of
  ${\cal O}(1\%)$. Positron counting, however, is mandatory only for the absolute
  cross section normalization.

\bibitem{Agostinelli:2002hh}
  S.~Agostinelli {\it et al.}  [GEANT4 Collaboration],
  Nucl.\ Instrum.\ Meth.\ A {\bf 506} (2003) 250.

\bibitem{Allison:2006ve}
  J.~Allison, K.~Amako, J.~Apostolakis, H.~Araujo, P.~A.~Dubois, M.~Asai, G.~Barrand and R.~Capra {\it et al.},
  IEEE Trans.\ Nucl.\ Sci.\  {\bf 53} (2006) 270.

\bibitem{adloff:2010hb}
  C.~Adloff {\it et al.}  [CALICE Collaboration],
  JINST {\bf 5} (2010) P05004.

\bibitem{cavallari} For a review see e.g. F.~Cavallari, PoS (EPS-HEP
  2013) 490 and references therein. Available at \texttt{http://pos.sissa.it/} 

\bibitem{Simon:2013zya}
  F.~Simon, C.~Soldner and L.~Weuste,
  JINST {\bf 8} (2013) P12001.

\bibitem{Cartiglia:2013haa}
  N.~Cartiglia, M.~Baselga, G.~Dellacasa, S.~Ely, V.~Fadeyev, Z.~Galloway, S.~Garbolino and F.~Marchetto {\it et al.},
  JINST {\bf 9} (2014) C02001.

\bibitem{ATLAS:2010aba} 
G. Aad et al., [ATLAS Collaboration],
  ATLAS-CONF-2010-007, ATLAS-COM-CONF-2010-007.

\bibitem{NA62_TDR}
F.~Hahn et al., [NA62 Collaboration], ``NA62 Technical Design Document'',
NA62-10-07, 2010. 

\bibitem{Abe:2007bi}
  T.~Abe {\it et al.}  [ISS Detector Working Group Collaboration],
  JINST {\bf 4} (2009) T05001.

\bibitem{Antonello:2014rwa}
  M.~Antonello {\it et al.}  [ICARUS Collaboration],
  JINST {\bf 9} (2014) P08003.

\bibitem{Rubbia:2014dva}
  C.~Rubbia,
  arXiv:1408.6431 [physics.ins-det].

\bibitem{Kopp:2006ky}
  S.~E.~Kopp,
  Phys.\ Rept.\  {\bf 439} (2007) 101.

\end{thebibliography}
\end{document}